\documentclass[aps,prl,superscriptaddress,showpacs,twocolumn,letterpaper]{revtex4}

 \usepackage{graphicx} 
\usepackage {bm}

\newcommand {\bisco} {Bi$_2$Sr$_{2-x}$La$_x$CuO$_{6+\delta}$}

\begin{document}

\title {Normal State Spectral Lineshapes of Nodal Quasiparticles in Single Layer Bi2201 Superconductor}

\author {A. Lanzara}
\affiliation {Department of Physics, University of California,
Berkeley, California 94720, USA}
\affiliation {Materials Sciences Division, Lawrence Berkley
National Laboratory, Berkeley, California 94720, USA}
\author {P. V. Bogdanov}
\author {X. J. Zhou}
\affiliation {Department of Physics, Applied Physics and Stanford
Synchrotron Radiation Laboratory, Stanford University, Stanford,
California 94305, USA}
\author {N. Kaneko}
\author {H. Eisaki}
\affiliation {Electrotechnical Laboratory, Tsukuba, Japan}
\author {M. Greven}
\affiliation {Department of Physics, Applied Physics and Stanford
Synchrotron Radiation Laboratory, Stanford University, Stanford,
California 94305, USA}
\author {Z. Hussain}
\affiliation {Advanced Light Source, Lawrence Berkeley National
Laboratory, Berkeley, California 94720, USA}
\author {Z. -X. Shen}
\affiliation {Department of Physics, Applied Physics and Stanford
Synchrotron Radiation Laboratory, Stanford University, Stanford,
California 94305, USA}
\date {\today}

\begin {abstract}
A detailed study of the normal state photoemission lineshapes and
quasiparticle dispersion for the single layer \bisco (Bi2201)
superconductor is presented. We report the first experimental
evidence of a double peak structure and a dip of spectral
intensity in the energy distribution curves (EDCs) along the nodal
direction. The double peak structure is well identified in the
normal state, up to ten times the critical temperature. As a
result of the same self-energy effect, a strong mass
renormalization of the quasiparticle dispersion, i.e. kink, and an
increase of the quasiparticle lifetime in the normal state are
also observed. Our results provide unambiguous evidence on the
existence of bosonic excitation in the normal state, and support a
picture where nodal quasiparticles are strongly coupled to the
lattice.
\end {abstract}

\pacs {74.25.Jb, 74.72.-h, 79.60.-i, 71.38.-k}
\maketitle


As a measure of the imaginary part of the single particle Green's
function, photoelectron spectroscopy provides insights on the
scattering processes that play a key role for the physical
properties of a material.  In a metallic system for example, the
coupling of quasiparticles to phonons causes the photoemission
line shape to evolve from a single Lorentzian-like peak, as for a
Fermi liquid picture, to double or multiple peak structures
\cite{one} with a dip of spectral intensity, which correspond to
the phonons energy. The coupling to phonon, or more in general to
any bosonic excitation, is also reflected in a renormalization of
the quasiparticle dispersion, and in an increase of the
quasiparticle lifetime below the phonon energy \cite{two}.  In the
framework of quasiparticles coupled to a bosonic excitation these
behaviors are the result of the same self-energy effect,
$\Sigma$($k$, $\omega$), and occur at a similar energy scale
\cite{three}.

This textbook behavior has been recently measured by angle
resolved photoemission spectroscopy (ARPES) in several systems
characterized by a strong electron-phonon interaction, such as Be
\cite{four}, Mo \cite{five}, W \cite{six} and $C_{60}$ \cite{seven}.
Similar behavior has also been observed in several families of
p-type cuprates along the nodal direction, (0, 0) to ($\pi$,
$\pi$) \cite{eight,nine,ten,eleven,twelve,thirteen,fourteen,fifteen}. While in the case of simple metals the interpretation is
straightforward and phonons are easily identified as the relevant
energy scale, the interpretation is far more complex in the case
of strongly correlated systems as cuprates superconductors.
Although the experimental data between different groups are in
agreement, and are in favor of a scenario of quasiparticles
coupled to bosonic modes, the nature of such excitations is still
highly controversial and has been matter of intense study over the
last few years. The two proposed scenarios see quasiparticles
coupled to phonons \cite{nine, ten} vs quasiparticles coupled to
an electronic mode \cite{eleven, twelve, thirteen, fourteen, fifteen}. On the basis of energy scale it has been argued that,
the highest energy phonon coupled to quasiparticle is the in plane
half-breathing oxygen phonons \cite{sixteen, seventeen}, while the
relevant electronic mode is the ($\pi$, $\pi$) resonance mode,
observed by neutron scattering \cite{eighteen, nineteen}. The
major difference between these two proposed scenario lies on their
normal state behavior. In the case of phonons a well-defined
energy scale in the normal and superconducting state is needed,
while, in the case of resonance mode scenario, no energy scale is
defined in the normal state, where the quasiparticle behave as
Marginal Fermi liquid \cite{twenty}. It has been argued that, the
Marginal Fermi liquid scenario can account for the following
normal state behaviors observed in Bi2212: curvature in the
dispersion \cite{eleven} that is related to the linear energy
dependence of the scattering rate \cite{twentyone}, and the
presence of an additional structure in the EDCs only in the
superconducting state \cite{eleven}. However there are two issues
that are subjects of continuous debate. First, the marginal Fermi
liquid picture itself does not define a microscopic origin so a
combined electronic and phononic contribution could give rise to a
behavior mimic the Marginal Fermi liquid phenomenology. Second,
and more substantively, the nodal kink is found to exist over the
entire doping range [10] with similar energy scale, while the
marginal Fermi liquid behavior is observed only near optimal
doping. Further, the kink in the dispersion is very sharp in
underdoped samples and not possible to be fit by a smooth band as
in the case of Marginal Fermi liquid.

An investigation of the normal state energy distribution curve
(EDC) can provide a definitive answer to this issue. In the
Marginal Fermi liquid scenario, the EDC curve is a single
lorentzian peak, where the peak width, proportional to the
imaginary part of the electron self-energy, depends linearly on
the frequency. This condition is dictated by the absence of any
energy scale other than the high frequency cut off in the Marginal
Fermi liquid picture.  In a phonon type scenario, on the other
hand, the phonon energy scale will show up in the EDC spectral
function, which is now characterized by a double peak structure,
the so-called peak-dip-hump structure \cite{one}.  This expected
contrast thus provides an acid test for the two cases.  This test
can be performed in the single layer Bi2201 system, as we can have
access to the normal state behavior without suffering from thermal
effects. In addition the single layer Bi2201 spectra do not suffer
from bilayer splitting effect, which might make the picture more
complex in the case of the double layer Bi2212.  Recently in fact,
an alternative explanation to the nodal double peak structure
observed in Bi2212 \cite{nine} has been proposed \cite{twentysix},
where the two peaks have been associated with the O$_2ps$-O$_2ps$
bands of two adjacent CuO layers \cite{twentysix}, i.e. bilayer
splitting, rather than as a signature of electrons coupled to a
collective mode \cite{one}. The existence of bilayer splitting
does not rule out coupling to collective mode, something made very
clear by the data from the antinode.  The presence of bilayer splitting does
make the data analysis more complex.

Here we report a detailed ARPES study of the normal state
quasiparticles in single layer underdoped Bi2201 ($T_c$= 10 K).
Our results provide the first evidence of a double peak structure
in the EDCs along the nodal direction, (0, 0) to ($\pi$, $\pi$).
The double peaks structure persists well above the critical
temperature, up to ten times $T_c$. A kink in the dispersion and
an increase of the quasiparticle lifetime at the dip energy are
also observed. These results provide unambiguous evidence of
quasiparticles coupled to bosonic excitations in the normal state,
and put a strong constraint on the fundamental scattering process
of cuprates, supporting a picture where phonons are strongly
involved.


Angle resolved photoemission data have been recorded at beam-line
10.0.1.1 of the Advanced Light Source in Berkeley, in a similar
set-up as reported previously \cite{twentythree}. The underdoped
(UD) \bisco (x=0.75, $T_c$= 10 K) was grown using the floating-zone
method. The data were collected utilizing 33eV photon energy. The
momentum resolution was $\pm 0.14$ degrees and the energy
resolution 14meV. The single crystalline samples were cleaved in
situ at low temperature and the samples were oriented so that the
analyzer scan spans along the (0, 0) to ($\pi$, $\pi$) diagonal
direction. We have performed a mapping at low temperature (20 K)
and a temperature dependent study along the nodal, (0, 0) to
($\pi$, $\pi$), direction. The temperature dependence was studied
through a thermal cycle in order to monitor the sample quality
during the experiment and the reproducibility of the observed
behavior. The vacuum during the measurement was better than
$\sim$ 4e-11 Torr. In this paper we use the notation low and
high energy to indicate energy range of (0, -60meV) and energy
range of (-60 to -300meV) respectively.
\begin {figure} [!t]
\includegraphics[width=2in]{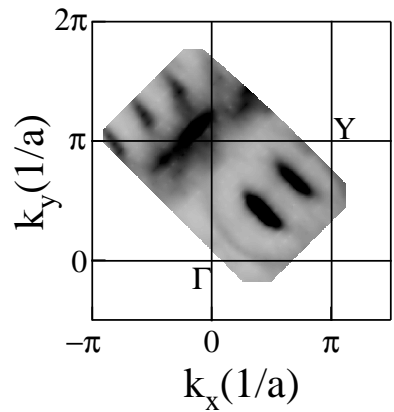}

\caption {Map of the ARPES intensity in the momentum space at the
Fermi energy for underdoped Bi2201 superconductors. The map is
taken at 25K. In the maps, black corresponds to the maximum
intensity and white to zero intensity.}

\end {figure}

In figure 1, we show the map of the spectral intensity at the Fermi
energy ($E_F$) in the normal state, up to the second Brillouin
zone (BZ). Highest intensity points in the spectral intensity map
at the Fermi energy (black color) give the Fermi surface. As
previously reported, the spectral weight primarily concentrates
around the nodes along the (0, 0)  to ($\pi$, $\pi$) direction of
the superconducting gap function \cite{twentyfour, twentyfive}.
The crossing between the superstructures bands and their replica
enhances the spectral weight at ($\pi$, 0). The contribution
coming from the main and the superstructure replica are clearly
seen and can be well separated in a momentum window close to the
nodal direction. This global mapping allows us to locate the
momentum precisely.

In figure 2, we show high-resolution energy distribution curve 
(EDCs), ARPES curves as a function of energy, along the nodal
direction at several temperatures, up to more than ten times
$T_c$. All the data reported here are in the normal state for
$T_c<T<T$*, where $T$* is the pseudogap temperature.  The EDCs
stack shows two main features: sharp slow-dispersing low binding
energy peak, crossing the Fermi energy $E_F$, defined to be zero
throughout this paper, and, broad fast-dispersing high binding
energy peak (hump). The two peaks coexist in a small momentum
interval and are separated by a dip of spectral intensity at
$\sim$60meV (dashed lines in the figure). Similar lineshapes have been
reported in the case of Be \cite{four} and for double layer Bi2212
in the superconducting state \cite{nine}. Finally, as the
temperature is raised (from 20K to 200K) the two peaks evolve in a
single broad feature and the dip between them is hardly
distinguishable. This is simply a thermal effect as explained
before and it is observed in Be too \cite{four}. In addition, the data
shown here, where a double structure can be distinguished despite
the absence of bilayer splitting, clearly rule out the
interpretation proposed to explain the Bi2212 double peak
structure in terms of bilayer splitting \cite{twentysix}. Our data
support therefore the following arguments: 1) Presence of a double
peak structure for nodal quasiparticles; 2) Persistence of the
double peak structure well above the critical temperature,
suggesting that bosons with well defined frequency are coupled to
quasiparticles in the normal state; 3) Temperature is the main
cause of the broadening observed in the EDCs at high temperature.
\begin {figure} [!t]

\includegraphics [width = 3.3 in]{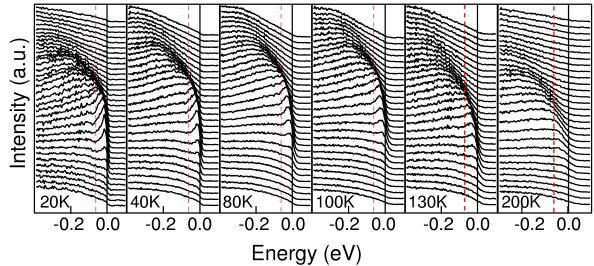}

\caption {Energy Distribution curves (EDC) in the normal state at
several temperatures (from 20K to 200K) for the underdoped Bi2201
superconductors. All the data here shown are in the normal state.
A dip in the EDCs can be clearly observed almost for all the
temperatures. The dip position (dotted line) is $\sim 60meV$
and is roughly temperature independent.}

\end {figure}


\begin {figure} [!b]

\includegraphics [width = 3.3 in]{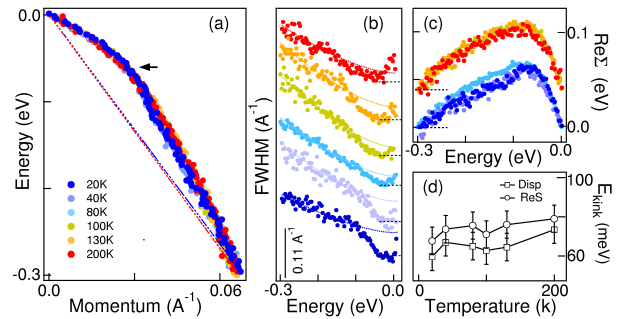}

\caption {a) Quasiparticle dispersions, extracted by fitting the
momentum distribution curves, along the nodal direction. The
dispersions are shown for same temperature range as in fig. 2. The
bare quasiparticle dispersion $\epsilon$(k) is shown by dotted
lines. Details of how to determine $\epsilon$(k) are given in the
text. b) Width of the momentum distribution curves (FWHM) vs
quasiparticle energy in the nodal direction, for underdoped
Bi2201. The FWHM at each temperature are shifted by a constant
offset (0.08$A^{-1}$). A step like function is observed for all
measured temperatures, however it becomes smoother as the
temperature increases. The drop of the MDC width corresponds to
the energy kink position. c) Re $\Sigma$ ($k$, $\omega$) vs
binding energy determined from the difference between the measured
quasiparticle dispersion and the bare dispersion (dashed line in
panel a). The Re $\Sigma$ ($k$, $\omega$)  at high temperature are
shifted by a constant offset (40meV) from the low temperature
data. d) Kink energy position vs temperature as extracted by two
different methods: 1) maximum position of the Re $\Sigma$ ($k$,
$\omega$)(circles); 2) straight lines fit of the low and high
energy quasiparticle dispersions (squares).}

\end {figure}

In figure 3a, we show the nodal quasiparticle dispersion vs binding
energy for several temperatures, same as shown in figure 2. The
dispersions are extracted by fitting the momentum distribution
curves (MDC), cut at fixed energy, with Lorentzian like functions.
A sharp break in the dispersion, kink, is observed (see horizontal
arrows in the figure), in agreement with previous results
\cite{nine, thirteen}. Consistently with the temperature
dependence of the ARPES spectra in figure 2, the kink in the
dispersion is well defined up to very high temperatures. At
temperatures high enough ($\sim 130K$) however, the kink
structure become smoother. A similar smoothening, in the case of
Bi2212, has been erroneously argued as evidence of no energy scale
\cite{eleven}. The data presented here in fact, clearly show that
the smoothening of the kink is simply an effect of temperature
broadening.

 In panel b, we report the MDC width at half maximum (FWHM) vs binding energy for same
 temperatures as in panel a. Each curve is shifted by a constant offset (see caption).
 Within the sudden approximation, the MDC width is proportional to the imaginary part
 of the electron self-energy, Im$\Sigma$($k$, $\omega$) \cite{three}.  The FWHM consists of two components: i)
 a step like sudden increase near $\sim 60meV$, as predicted in the case of quasiparticles
 coupled to a sharp bosonic excitation; ii) a $\omega^2$ dependence that persists up to very
 high energy.  The step like structure in the FWHM is defined for all temperatures.
 In the hypothesis of independent scattering mechanism, valid in weak coupling regime \cite{four, five},
 the step like increase at low energy in our data can be interpreted as due to coupling to bosons.
 Therefore, the presence of a step like structure in the normal state is a clear evidence that
 similar coupling persists up to very high temperatures. This observation is also consistent
 with the temperature dependence of the kink in the dispersion (fig. 3a) and the temperature
 dependence of the double peaks structure in the ARPES spectra (fig. 2). Similar behavior of
 the FWHM in the normal state has been reported for the La$_{2-x}$Sr$_x$CuO$_4$ (LSCO) compound at x=0.03
 doping \cite{ten}.  Finally, the data here presented, clearly support a picture where coupling to
 the same bosonic modes is present in the normal state.  In panel c we show the real part of
 the electron self energy, Re $\Sigma$($k$, $\omega$), vs binding energy in the same temperature range as in
 previous panels. The Re$\Sigma$($k$, $\omega$)  is defined as the difference between the measured (dressed)
 quasiparticle dispersion, and the bare dispersion, $\epsilon$(k) \cite{twentytwo}. Here we approximate the bare
 dispersion by a straight line connecting $k_F$ and the dispersion at high energy 300meV, see dotted
 lines in panel a), following a previous convention \cite{eleven}.  The value of the Re$\Sigma$($k$, $\omega$)  is of
 course strongly dependent from our choice of $\epsilon$(k), however this type of analysis hold for a
 qualitative comparison between different temperatures, since the bare dispersion should be
 temperature independent.  The maximum position of the Re$\Sigma$($k$, $\omega$)  is therefore another way to
 identify and measure the kink position in the dispersion. At low temperatures a well-defined
 peak in the Re$\Sigma$($k$, $\omega$) is observed.

The peak broadens as the temperature increases as a result of
thermal broadening, as discussed in panel a and b. The broadening
of the Re$\Sigma$($k$, $\omega$)  can also be predicted within an
Eliashberg type electron-boson model, see ref. \cite{twentytwo}.
It is important to point out that, in the case of Bi2212, the
broadening of the Re$\Sigma$($k$, $\omega$) and the disappearance
of the sharp peak at low energy, have been considered compelling
evidence against a electron-boson interacting picture, where the
broad hump in the Re$\Sigma$($k$, $\omega$) is now purely
dominated by electron-electron interaction \cite{eleven,
twentyseven}. The data here shown, where a sharp peak in the
Re$\Sigma$($k$, $\omega$) is observed well above $T_c$ suggest
that temperature effect might be the cause of the broadening
observed in the Bi2212 case, where $T_c$ is higher \cite{eleven,
twentyseven}.

In panel d we report the energy position of the nodal kink as a
function of temperature. The kink position is extracted using two
correlated methods: 1) the maximum position of the Re$\Sigma$($k$,
$\omega$) (circles) \cite{twentytwo}; 2) the energy intersection
between the straight line fit for the low and high energy part of
the dispersion \cite{nine} (squares). The data show a slight
temperature dependent increase of the kink position, as explained
in simple electron-boson picture \cite{twentytwo}. The large error
bars at high temperatures are due to the broader peaks with
respect to the low temperature data.

In conclusion, we have shown a detailed study of the photoemission
spectra in the underdoped single layer Bi2201 superconductor. We
have provided the first evidence of a double peak structure in the
nodal ARPES spectra up to ten times the critical temperature. The
dip of the spectral weight corresponds to the same energy scale
where the quasiparticle dispersion shows a kink, or similarly
where Re$\Sigma$($k$, $\omega$)shows a peak, and where the
Im$\Sigma$($k$, $\omega$) shows a step like change. All the
results here presented clearly support a scenario of
quasiparticles coupled to bosonic excitations with well defined
energy in the normal state. Given the absence of a resonance mode
in the single layer Bi2201 and the temperature dependence behavior shown here, we conclude that the bosonic excitations are
associated with coupling with oxygen related optical phonons.

\begin{acknowledgments}

This work was supported by the Department of Energy's Office and
Basic Energy Science, Division of Materials Sciences and
Engineering of the U.S. Department of Energy, under Contract No.
DE-AC03-76SF00098 and by the National Science Foundation through
Grant DMR-0349361. SSRL work is supported by DOE contract
DE-ACO3-765F00515. Stanford work is also supported by NSF
DMR-0304981.
\end{acknowledgments}


\begin{thebibliography}{99}

\bibitem{bogdanov} P. V. Bogdanov, Phys. Rev. Lett. 85, 2581 (2000)
\bibitem{one} D. J. Scalapino in Superconductivity (ed. Parks, R. D.) p. 449
\bibitem{two} F. Marsiglio Phys. Rev. B 47, 5419 (1993)
\bibitem{three} L. Hedin and S. Lundqvist in Solid State Physics, edited by Seitz, F., Turnbull,
D.\& Ehrenreich, H., Academic Press (1969).
\bibitem{four} M. Hengsberger et al., Phys. Rev. Lett. 83, 592 (1999); S. LaShell, E. Jensen, and
T. Balasubramanian, Phys. Rev. B 61, 2371 (2000)
\bibitem{five} T. Valla et
al. Phys. Rev. Lett. 83, 2085(1999)
\bibitem{six} E. Rotenberg et al. Phys. Rev. Lett. 84, 2925 (2000)
\bibitem{seven} O. Gunnarsson et al. Phys. Rev. Lett.74, 1875 (1995)
\bibitem{eight} P. V. Bogdanov et al. Phys. Rev. Lett. 85, 2581 (2000)
\bibitem{nine}  A. Lanzara et al. Nature 412, 510-514 (2001)
\bibitem{ten}  X. J. Zhou et al. Nature 423, 398 (2003)
\bibitem{eleven}  P. D. Johnson et al. Phys. Rev. Lett. 87, 177007
(2001)
\bibitem{twelve} A. Kaminski et al. Phys. Rev. Lett. 86, 1070 (2001)
\bibitem{thirteen}T. Sato et al. Phys. Rev. Lett. 91, 157003 (2003)
\bibitem{fourteen}A. D. Gromko et al. Cond-mat/0202329 (2003)
\bibitem{fifteen}T. K. Kim et al. Cond-Mat 0303422 (2003)
\bibitem{sixteen} R. J. McQueeney et al. Phys. Rev. Lett. 82, 628 (1999)
\bibitem{seventeen} L. Pintschovious et al. Phys. Rev. B. 60, 15039 (1999)
\bibitem{eighteen} H. F. Fong et al. Phys. Rev. B 54, 6708 (1996)
\bibitem{nineteen}  H. He et al. Cond-mat/0002013 (2000)
\bibitem{twenty}  C. M. Varma et al., Phys. Rev. Lett. 63, 1996 (1989)
\bibitem{twentyone}  A. Kaminski et al., Phys. Rev. Lett. 84, 1788 (2000)
\bibitem{twentytwo}  S. Verga et al. Cond-mat/0207145 (2003); to
be published in Phys. Rev. B.
\bibitem{twentythree}  P. V. Bogdanov et al. Phys. Rev. Lett. 89, 167002 (2002)
\bibitem{twentyfour}  D. S. Marshall et al. Phys. Rev. Lett. 76, 4841-4844 (1996)
\bibitem{twentyfive}   Norman, M.R. et al. Nature 392, 157 (1998)
\bibitem{twentysix}   A. A. Kordyuk et al. Cond-mat/0311137 (2003)
\bibitem{twentyseven}   H. Wang et al. Nature 427, 714 (2004)
\bibitem{twentyeight}  G. H. Gweon et al. J. Phys. and Chem. Solids 65, 1397 (2004)
\bibitem{twentynine}   T. Cuk et al. Phys. Rev. Lett. 93, 117003 (2004)
\bibitem{thirty} G. H. Gweon et al. Nature 430, 187 (2004)

\end{thebibliography}
\end{document}